\documentclass[aps,prb,reprint,twoside,showpacs]{revtex4-1}
\usepackage{amssymb}
\usepackage{amsmath}
\usepackage{graphicx}
\usepackage{CJK}
\usepackage[sf]{subfigure}
\usepackage[breaklinks=true,colorlinks=true,linkcolor=blue,citecolor=blue,urlcolor=black,bookmarks=false]{hyperref}
\usepackage{mathptmx}

\input{tcilatex}
\begin{document}

\begin{CJK*}{Bg5}{bsmi}
\title{Spin-dependent Klein tunneling in graphene: Role of Rashba spin-orbit coupling}
\author{Ming-Hao Liu (¼B©ú»¨)}
\affiliation{Institut f\"{u}r Theoretische Physik, Universit\"{a}t Regensburg, D-93040 Regensburg, Germany}%
\author{Jan Bundesmann}
\affiliation{Institut f\"{u}r Theoretische Physik, Universit\"{a}t Regensburg, D-93040 Regensburg, Germany}%
\author{Klaus Richter}
\affiliation{Institut f\"{u}r Theoretische Physik, Universit\"{a}t Regensburg, D-93040 Regensburg, Germany}%
\begin{abstract}%
Within an effective Dirac theory the low-energy dispersions of monolayer
graphene in the presence of Rashba spin-orbit coupling and spin-degenerate
bilayer graphene are described by formally identical expressions.
We explore implications of this correspondence for transport by
choosing chiral tunneling through \textit{pn }and \textit{pnp}
junctions as a concrete example. A real-space Green's function formalism
based on a tight-binding model is adopted to perform the ballistic
transport calculations, which cover and confirm previous theoretical results based on
the Dirac theory. Chiral tunneling in monolayer graphene in the presence of
Rashba coupling is shown to indeed behave like in bilayer graphene.
Combined effects of a forbidden normal transmission and spin separation
are observed within the single-band $n\leftrightarrow p$ transmission
regime. The former comes from real-spin conservation, in analogy with
pseudospin conservation in bilayer graphene, while the latter arises from the
intrinsic spin-Hall mechanism of the Rashba coupling.
\end{abstract}
\date{\today}
\maketitle
\end{CJK*}

\section{Introduction\label{sec introduction}}

After the first successful isolation of monolayer graphene (MLG) was
announced,\cite{Novoselov2004} intriguing properties based on its low-energy
excitation that mimics massless, gapless, and chiral Dirac fermions were
intensively investigated.\cite{CastroNeto2009,DasSarma2011} Spin-orbit
coupling (SOC), on the other hand, is the key ingredient of semiconductor
spintronics\cite{Awschalom2002,Zutici2004} that was undergoing a rapid
development before the rise of graphene.\cite{Geim2007} The question about
the role of SOC effects in graphene then naturally emerged, including the
proposal of graphene as a topological insulator,\cite{Kane2005} which
attracted the attention of various first-principles-based studies.\cite%
{Min2006,Gmitra2009,Abdelouahed2010}

SOC in MLG includes an intrinsic and an extrinsic term. The former reflects
the inherent asymmetry of electron hopping between next nearest neighbors%
\cite{Kane2005} (i.e., a generalization of Haldane's model\cite{Haldane1988}).
The latter is induced by the electric field perpendicular to the graphene
plane, which can be externally controlled, and resembles the Rashba model%
\cite{Rashba1960,Bychkov1984} for the two-dimensional electron gas.
Agreement has been achieved, based on first-principles calculations,\cite%
{Gmitra2009,Abdelouahed2010} that the intrinsic SOC term opens a gap of the
order of $2\lambda _{I}\approx 24%
\unit{\mu eV}%
$, while the Rashba SOC removes the spin degeneracy and creates a
spin-splitting $2\lambda _{R}$ at the $K$ and $K^{\prime }$ points that has
a linear dependence on an external electric field $E$ with the slope of
about $100%
\unit{\mu eV}%
$ per $\unit{V}/\unit{%
\text{\AA}%
}$ of $E$. Under a strong gate voltage, the Rashba coupling may in principle
dominate the intrinsic SOC in MLG.\cite{Gmitra2009,Abdelouahed2010}

\begin{figure}[b]
\includegraphics[width=\columnwidth]{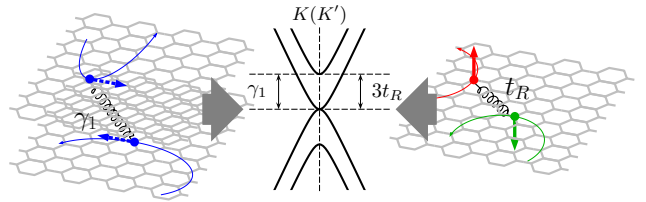}
\caption{(Color online) Schematic of the pseudospin coupling through $%
\protect\gamma _{1}$ in BLG (left panel) and real-spin coupling through $%
t_{R}$ in MLG (right), which lead to an identical low-energy dispersion near
$K$ and $K^{\prime }$.\label{fig1}}
\end{figure}

The low-energy spectrum of MLG plus the Rashba coupling (MLG+R) was derived
by Rashba,\cite{Rashba2009} based on the Kane-Mele model\cite{Kane2005} (
i.e., an effective Dirac Hamiltonian). An earlier work by one of us\cite%
{Liu2009} started with a tight-binding model (TBM) and obtained an
equivalent form of the low-energy expansion,\footnote{%
Due to a minor difference in the definition of the Rashba coupling in the
tight-binding Hamiltonian, the splitting $3t_{R}$ here corresponds, e.g., to
$\lambda $ in Ref.\ \onlinecite{Rashba2009} and to $2\lambda _{R}$ in Ref.\ %
\onlinecite{Gmitra2009}.}%
\begin{equation}
E_{\text{MLG+R}}\left( \mathbf{q}\right) \approx \mu \tfrac{1}{2}[\sqrt{%
\left( 3t_{R}\right) ^{2}+\left( 3ta\cdot q\right) ^{2}}+\nu \left(
3t_{R}\right)],  \label{MLG+R}
\end{equation}%
which also agrees with expressions given in Refs.\ \onlinecite{Gmitra2009}
and \onlinecite{Abdelouahed2010} when $\lambda _{I}=0$. Here $\mu ,\nu =\pm
1 $ are band indices, $t$ and $t_{R}$ are nearest-neighbor kinetic and
Rashba hopping parameters, respectively, $a\approx 1.42\unit{%
\text{\AA}%
}$ is the bonding length, and $\mathbf{q}=K+\delta \mathbf{k}$ with $%
\left\vert \delta \mathbf{k}\right\vert a\ll 1$. Recall for comparison the
low-energy spectrum of bilayer graphene (BLG),\cite%
{McCann2006,CastroNeto2009}
\begin{equation}
E_{\text{BLG}}\left( \mathbf{q}\right) \approx \mu \tfrac{1}{2}(\sqrt{%
\gamma _{1}^{2}+\left( 3ta\cdot q\right) ^{2}}+\nu \gamma _{1}),
\label{BLG}
\end{equation}%
where $\gamma _{1}$ is the nearest-neighbor hopping between the two graphene
layers. Note that the next nearest-neighbor interlayer hoppings $\gamma _{3}$
and $\gamma _{4}$ do not influence the band dispersion near $K$. The
completely different mechanisms of (i) pseudospin coupling between carriers
from the two graphene layers of BLG through interlayer hopping $\gamma _{1}$
and (ii) real-spin coupling between up and down spins within MLG through
Rashba hopping $t_{R}$ happen to lead to an identical mathematical form in
Eqs.\ \eqref{MLG+R} and \eqref{BLG} that can be clearly mapped onto each
other\cite{Yamakage2009,Rakyta2010} with $\gamma _{1}\leftrightarrow 3t_{R}$
as sketched in Fig.\ \ref{fig1}. This unambiguously implies that low-energy
physics in MLG+R and BLG should behave similarly.

In this paper we tackle the question of whether the transport in MLG+R
behaves as in BLG by choosing the issue of Klein tunneling\cite%
{Beenakker2008,CastroNeto2009,DasSarma2011,Allain2011} (or, in general,
chiral tunneling) as a concrete example. Chiral tunneling in graphene has
been shown to exhibit completely different behavior in MLG and BLG based on
the Dirac theory.\cite{Katsnelson2006} Tunneling at normal incidence in MLG
shows a suppression of backscattering, which resembles the original Klein
paradox in relativistic quantum electrodynamics\cite{Klein1929} and hence
the name Klein tunneling, while in BLG it shows a perfect reflection, which
is strictly speaking a consequence of forbidden interband transition also
due to the chiral nature of graphene. The theoretical discussion of chiral
tunneling so far focuses mainly on spin-independent tunneling through
\textit{pn }and \textit{pnp} junctions,\cite%
{Katsnelson2006,Cheianov2006,Zhang2008,Sonin2009,Bai2010a,Rossi2010,Pereira2010,Allain2011,Tudorovskiy2011}
while SOC effects are less discussed.\cite%
{Yamakage2009,Bercioux2010,Bai2010,Rataj2011} In addition, the relevant
theoretical understanding so far is based on Dirac theory, which is valid
only for the Fermi level close to the charge neutrality point and allows
only to consider certain relatively simple systems. A recent study
discussing the interplay between the Aharanov-Bohm effect and Klein
tunneling in graphene, started with a TBM,\cite{Schelter2010} but the
nanoribbon type of the leads used in that work may have edge effects
included that can be very different from the bulk properties of graphene. A
more transparent theoretical study of chiral tunneling in graphene directly
bridging the analytical Dirac theory and the numerical TBM computation is so
far missing and deserves consideration.

In the present work, we re-treat this issue of chiral tunneling in graphene
based on the TBM and show a unified description, allowing for a broad range
of geometries and complementing the existing results based on the Dirac
theory. Straightforward generalization to the case of MLG+R reveals a
spin-dependent tunneling behavior in close analogy with that in BLG, with
the role of pseudospin in BLG replaced by real spin in MLG+R. Specifically,
a combined behavior of spin-Hall-based spin separation and suppression of
normal transmission will be shown.

This paper is organized as follows. In Sec.\ \ref{sec formalism} we briefly
summarize the theoretical formalism applied in the present calculation,
namely, real-space Green's function formalism in noninteracting bulk
graphene. In Sec.\ \ref{sec results} we show our TBM results including the
consistency with the Dirac theory, a direct comparison between BLG and
MLG+R, and a deeper discussion of the MLG+R case. We review also briefly the
recent experimental progress on the Rashba spin splitting and Klein
tunneling in graphene in Sec.\ \ref{sec exp}, and finally conclude in
Sec.~\ref{sec conclusion}.

\section{Formalism\label{sec formalism}}

\subsection{Tight-binding model for \textquotedblleft
bulk\textquotedblright\ graphene\label{sec tbm for bulk}}

We choose the TBM for describing the electronic properties of graphene,
which is a well established way to treat graphene numerically. For
spin-degenerate MLG, the Hamiltonian reads

\begin{equation}
\mathcal{H}_{\text{MLG}}=\sum_{i}V_{i}c_{i}^{\dag }c_{i}-t\sum_{\langle
i,j\rangle }c_{i}^{\dag }c_{j},  \label{H MLG}
\end{equation}%
where the operator $c_{i}^{\dag }$ ($c_{i}$) creates (annihilates) an
electron at site $i$ (including both sublattices $A$ and $B$). The first sum
in Eq.\ \eqref{H MLG} runs over all the atomic sites in the considered
region with on-site potential $V_{i}$, and the second sum runs over all the
pairs of neighboring atomic orbitals $\langle i,j\rangle $ with kinetic
hopping parameter $t$ ($\approx 3\unit{eV}$). The next nearest neighbor
kinetic hopping term, usually characterized by $t^{\prime }\approx 0.1t$,
can be added in Eq.\ \eqref{H MLG} but will not be considered in the present
work due to the minor role it plays in the bulk transport properties for
low-energy excitation.

Spin-orbit interactions can be incorporated into the TBM by altering the
spin-dependent hopping between nearest and next-nearest neighbors,\cite%
{Kane2005,Konschuh2010} modifying Eq.\ \eqref{H MLG} as%
\begin{equation}
\mathcal{H}_{\text{MLG+R}}=\sum_{i}V_{i}\sigma ^{0}c_{i}^{\dag
}c_{i}+\sum_{\langle i,j\rangle }c_{i}^{\dag }\left[ -t\sigma ^{0}+it_{R}(%
\vec{\sigma}\times \mathbf{d}_{ij})_{z}\right] c_{j}.  \label{H MLG+R}
\end{equation}%
Here $\sigma ^{0}$ is the $2\times 2$ identity matrix, $t_{R}$ is the Rashba
spin-orbit hopping parameter, $\mathbf{d}_{ij}$ is the unit vector pointing
from site $j$ to $i$, and $\vec{\sigma}=\left( \sigma ^{x},\sigma
^{y},\sigma ^{z}\right) $ is the vector of (real-) spin Pauli matrices. We
take into account only the extrinsic SOC and neglect the intrinsic term in
order to highlight the role of the Rashba SOC.

For spin-degenerate BLG, we consider%
\begin{equation}
\mathcal{H}_{\text{BLG}}=\sum_{m=1,2}\mathcal{H}_{\text{MLG}}^{(m)}-\gamma
_{1}\sum_{j}\left( b_{2,j}^{\dag }a_{1,j}+\text{H.c.}\right) ,  \label{H BLG}
\end{equation}%
where $\mathcal{H}_{\text{MLG}}^{(m)}$ is $\mathcal{H}_{\text{MLG}}$ given
by Eq.\ \eqref{H MLG} of the $m$th graphene layer, $a_{m,j}$ ($b_{m,j}$)
annihilates an electron on sublattice $A$ ($B$) in layer $m=1,2$ at unit
cell $j$ (that contains two sublattice sites belonging to $A$ and $B$), and
the interlayer coupling strength $\gamma _{1}\approx 0.4\unit{eV}$
corresponds to the nearest neighbor hopping between the two MLG layers.
Further interlayer hopping terms,\cite{CastroNeto2009} $-\gamma
_{4}\sum_{j}(a_{2,j}^{\dag }a_{1,j}+b_{2,j}^{\dag }b_{1,j}+$ H.c.$)$ and $%
-\gamma _{3}\sum_{j}(a_{2,j}^{\dag }b_{1,j}+$ H.c.$)$, are not considered in
the present calculation, since they do not influence the low-energy
excitation. Throughout the presentation of the numerical results in Sec.\ %
\ref{sec results}, the kinetic hopping parameters will be fixed at $t=3\unit{%
eV}$ and $\gamma _{1}=0.39\unit{eV}$, while the value of the Rashba hopping
parameter $t_{R}$ depends on the context.

\begin{figure}[b]
\includegraphics[width=\columnwidth]{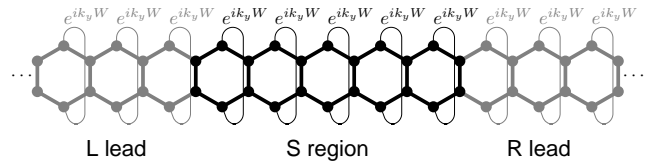}
\caption{Schematic of a minimum tight-binding model that simulates a bulk
MLG up to nearest neighbor hoppings with $W=3a$. Further nearest neighbor
hoppings can be accounted for by enlarging the transverse periodicity $W$ to
at least $6a$.\label{fig GNRwithPBC}}
\end{figure}

For the simulation of bulk graphene, we impose the Bloch theorem along the
transverse direction with periodicity $W$. This is equivalent to considering a
nanoribbon and modifying the hopping between atomic sites connected through the
periodic boundary conditions by a Bloch phase factor $e^{ik_{B}W}$ with a
Bloch momentum $k_{B}$,\cite{Wimmer2008} as schematically shown for MLG in
Fig.\ \ref{fig GNRwithPBC}. At the same time the Bloch momentum is the
component of the electron's momentum perpendicular to the nanoribbon, hence
defining the propagation angle $\phi =\sin ^{-1}(k_{B}/k_{F})$, where $k_{F}$
is the Fermi wave vector. To be consistent with the literature related to
Klein tunneling based on the Dirac theory, in Sec.\ \ref{sec results} we
will refer to the Bloch momentum as $k_{y}$.

In the present calculations, we will apply a minimal TBM by imposing the
periodic boundary conditions on a zigzag nanoribbon with chain number $%
N_{z}=2$, that is, periodicity of $W=3a$ (as the case sketched in Fig.\ \ref%
{fig GNRwithPBC}). The present model applies equally well for\ metallic
armchair ribbon (chain number $N_{a}$ being a multiple of $3$) with periodic
boundary conditions, but the minimal model would require $N_{a}=3$ (i.e.,
periodicity of $W=3\sqrt{3}a$).

\subsection{Brief summary of real-space Green's function formalism}

We consider open systems connected to the outer world by two leads (see
Fig.\ \ref{fig GNRwithPBC}). According to the real-space Green's function
formalism\cite{Datta1995} we numerically calculate the Green's functions of
our system,%
\begin{equation}
G_{S}^{r/a}=[E-H_{S}-\Sigma ^{r/a}\pm i\eta ]^{-1},  \label{Gr Ga}
\end{equation}%
where the self-energies of the leads ($\Sigma ^{r/a}=\Sigma
_{L}^{r/a}+\Sigma _{R}^{r/a}$) reflect the fact that our system is open. The
powerful recipe constructed in Ref.\ \onlinecite{Wimmer2008} for graphene
handles a lead as a semi-infinite repetition of unit cells and allows for
incorporating any kind of lattice structure and one-body interaction such as
SOCs. The transmission probability for an electron traveling from lead $L$
to lead $R$ is given by the Fisher-Lee relation\cite{Datta1995,Wimmer2008}%
\begin{equation}
T_{RL}=\func{Tr}(\Gamma _{L}G_{S}^{r}\Gamma _{R}F_{S}^{a}),  \label{T}
\end{equation}%
where the trace is done with respect to the lattice sites. The spectral
matrix functions $\Gamma _{L/R}$ are given by the lead self-energies as $%
\Gamma _{L/R}=i(\Sigma _{L/R}^{r}-\Sigma _{L/R}^{a})$.

For a given Bloch momentum $k_{y}$ and a given Fermi energy $E_{F}$ [subject
to a Fermi wave vector $k_{F}$ via Eq.\ \eqref{MLG+R} for MLG+R or Eq.\ %
\eqref{BLG} for BLG], the incoming propagation angle $\phi $ of the electron
wave can be defined as $\phi =\sin ^{-1}(k_{y}/k_{F})$. The angle-dependent
transmission function $T\left( \phi \right) $ is obtained from\ Eq.\ %
\eqref{T}, which can be generalized to a spin-resolved version.\cite%
{Nikolic2009}

\section{Transport results\label{sec results}}

In this section we present numerical results of our tight-binding transport
calculations. We first show the consistency of our tight-binding
calculations with the existing effective Dirac theory in Sec.\ \ref{sec
results part 1}. A direct comparison between BLG and MLG+R will then be
shown in Sec.\ \ref{sec results part 2}. Finally, Sec.\ \ref{sec results
part 3} is devoted to MLG+R for \textit{pn }junctions, in particular the
role of Rashba SOC for chiral tunneling.

\subsection{Consistency with Dirac theory\label{sec results part 1}}

We first consider tunneling in graphene without SOC and confirm existing
results, limited to low energy excitations, by our tight-binding
calculations. We pick two pioneering theoretical works to demonstrate the
consistency explicitly. Consistency with recent works of tunneling in
graphene heterojunctions in the presence of SOC\cite%
{Yamakage2009,Bercioux2010} has also been checked, but is not explicitly
shown here.

\begin{figure}[t]
\subfigure[]{
   \includegraphics[width=0.5\columnwidth]{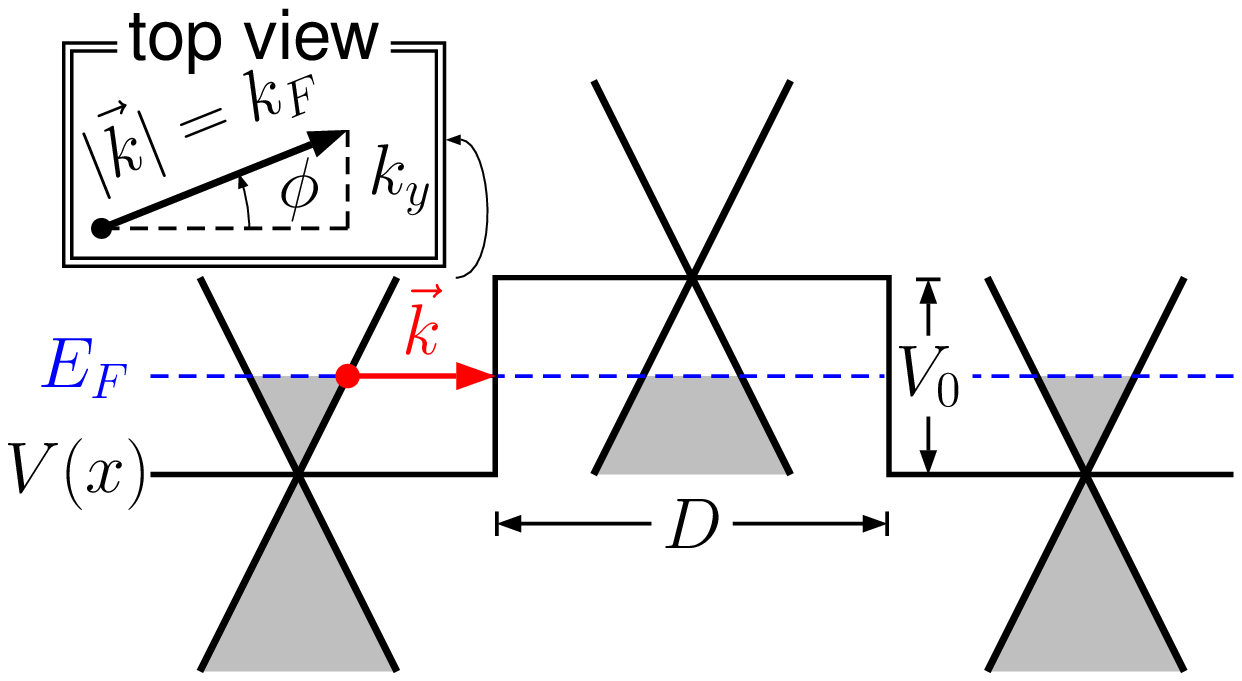}
   \label{fig_pnpMLG}}%
\subfigure[]{
   \includegraphics[width=0.5\columnwidth]{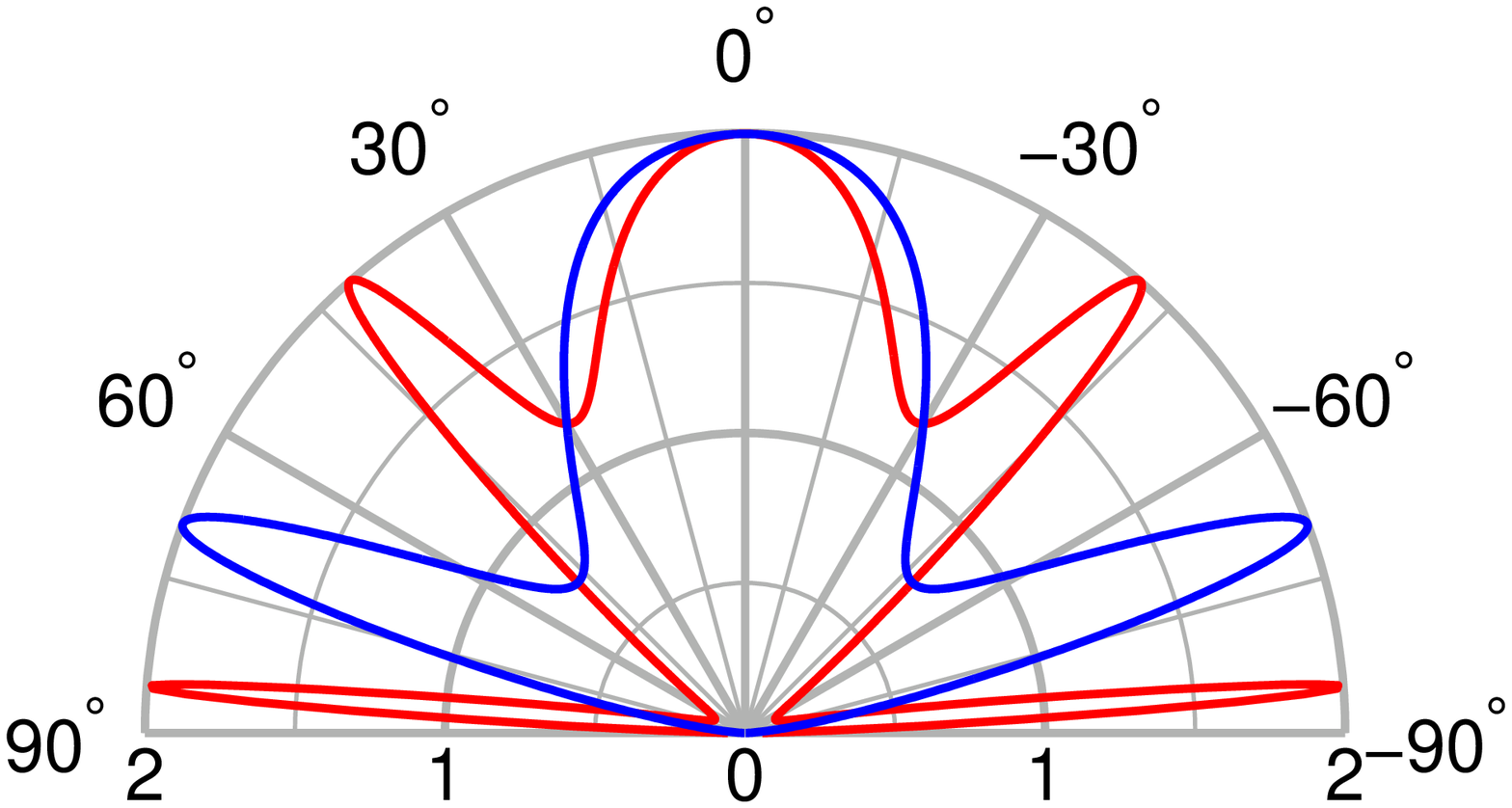}
   \label{fig_pnpMLG_T}}
\subfigure[]{
   \includegraphics[width=0.5\columnwidth]{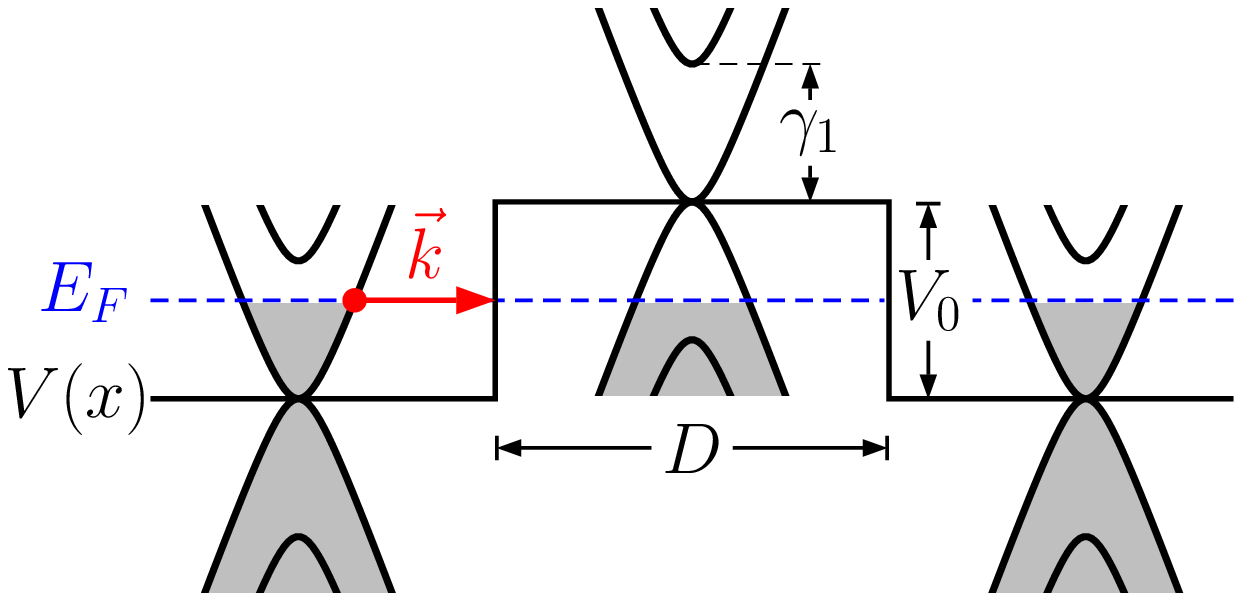}
   \label{fig_pnpBLG}}%
\subfigure[]{
   \includegraphics[width=0.5\columnwidth]{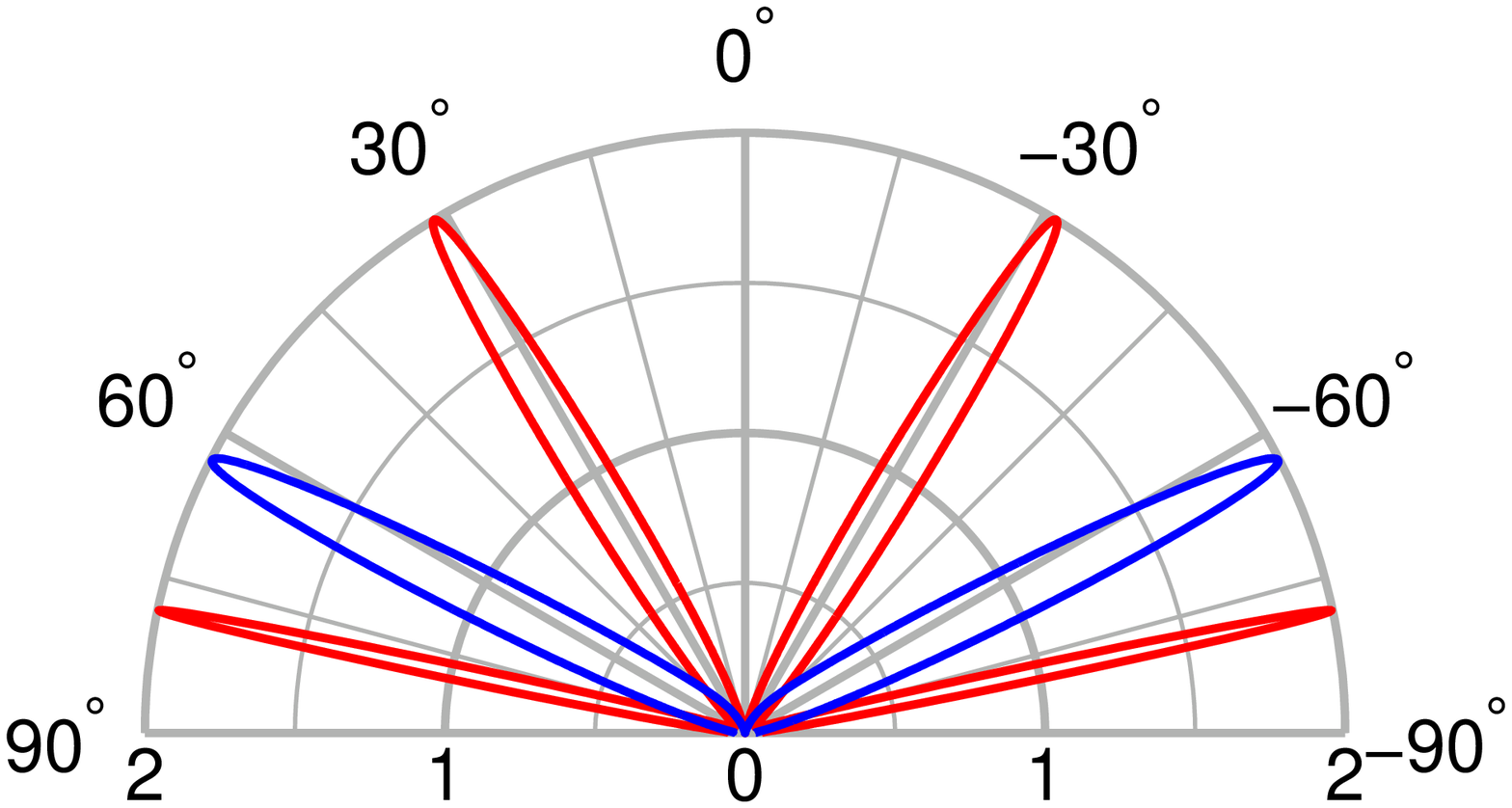}
   \label{fig_pnpBLG_T}}
\caption{(Color online) Tunneling through a barrier for (a), (b) MLG with $%
E_{F}=3tak_{F}/2=81.6\unit{meV}$ and (c), (d) BLG with $%
E_{F}=(3tak_{F}/2)^{2}/\protect\gamma _{1}=17.1\unit{meV}$. In (b), red
(light gray) and blue (dark gray) curves correspond to $V_{0}=196.8\unit{meV}
$ and $V_{0}=280.3\unit{meV}$, respectively. In (d), red (light gray) and
blue (dark gray) curves correspond to $V_{0}=48.7\unit{meV}$ and $V_{0}=100.7%
\unit{meV}$, respectively. In both cases the barrier width is $D=100\unit{nm}
$ and the incoming Fermi wave vector is $k_{F}=2\protect\pi /50\unit{nm}^{-1}$,
as considered in Ref.\ \onlinecite{Katsnelson2006}.\label{figMLGvsBLG}}
\end{figure}

\subsubsection{Chiral tunneling in MLG vs BLG}

Tunneling in MLG and BLG behaves quite differently as mentioned in Sec.\ \ref%
{sec introduction} and pointed out by Katsnelson \textit{et al.}\cite%
{Katsnelson2006} For a quantitative comparison we consider a barrier of
width $D=100\unit{nm}$ and the incoming Fermi wave vector $k_{F}=2\pi /50%
\unit{nm}^{-1}$ as in Ref.\ \onlinecite{Katsnelson2006} for both MLG and BLG [see
Figs.\ \ref{fig_pnpMLG} and \ref{fig_pnpBLG}]. Note that in order to exactly
match the barrier width, we set the bonding length $a=(4\sqrt{3})^{-1}\unit{%
nm}$, which differs from the realistic value of about $1.42\unit{%
\text{\AA}%
}$ by only less than $2\%$, so that the number of hexagons used here amounts
to $D/(\sqrt{3}a)=4\times \lbrack D]_{\unit{nm}}=400$.

The resulting transmission probabilities as a function of the incident angle
$\phi $ are depicted in Figs.\ \ref{fig_pnpMLG_T} and \ref{fig_pnpBLG_T}.
They reproduce the results of Fig.\ 2 in Ref.\ \onlinecite{Katsnelson2006}
almost perfectly, if we choose slightly different $E_{F}$ and $V_{0},$ to which
the transmissions at finite angles are sensitive. The remaining tiny
difference between our TBM results and their Dirac theory results\footnote{%
The unity transmission peaks (except the $0^{\circ }$ peaks for MLG) are
shifted by less than $3^{\circ }$ compared to Fig.\ 2 of Ref.\ %
\onlinecite{Katsnelson2006}.} simply reflects the basic difference between
the two approaches: For graphene the effective Dirac theory is valid only
for energies close to the Dirac point, while the TBM is suitable for the
entire energy range.

Note that the maximal values of the transmission functions in Fig.\ \ref%
{figMLGvsBLG} are $2$, since the valley degeneracy is automatically
incorporated in the tight-binding formalism. Later when we take spin also
into account, the maximum of the transmission function will be $4$. The
transmission probabilities calculated by the Dirac theory always have their
maximum of $1$ due to the normalized incoming wave, unless a proper
degeneracy factor is taken into account.

\subsubsection{Klein tunneling in MLG: Sharp vs smooth interface \label{sec
falko}}

Tunneling in MLG through a \textit{pn} junction exhibits probability one at
normal incidence and is called Klein tunneling. In experiments, a graphene
\textit{pn }junction can be realized by using a backgate, which tunes the
carrier density (and hence the Fermi level) globally, and a topgate that
tunes locally the carrier density, equivalent to the potential step $V_{0}$
at the other side.\cite{Williams2007} The carrier densities on the two sides
can be controlled to be of opposite signs, forming the \textit{pn} junction.
In between, however, the variation of the carrier density is never abrupt in
reality. Cheianov and Fal'ko showed, based on the Dirac theory, that the
interface of the \textit{pn} junction actually matters.\cite{Cheianov2006}
They considered symmetric \textit{pn} junctions (i.e., $V_{0}=2E_{F}$) with
sharp and linearly smooth interfaces, which we briefly review and compare
with our TBM results in the following.
\begin{figure}[t]
\subfigure[]{
   \hspace{-0.05cm}\includegraphics[width=0.5\columnwidth]{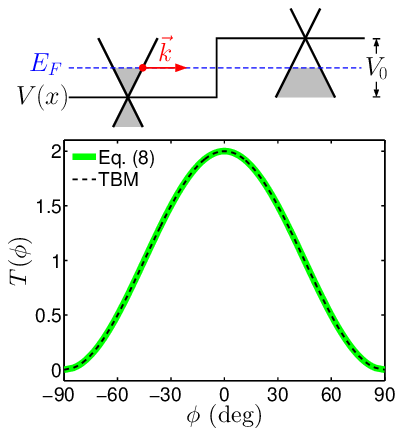}
   \label{fig sharp}}%
\subfigure[]{
   \hspace{-0.05cm}\includegraphics[width=0.5\columnwidth]{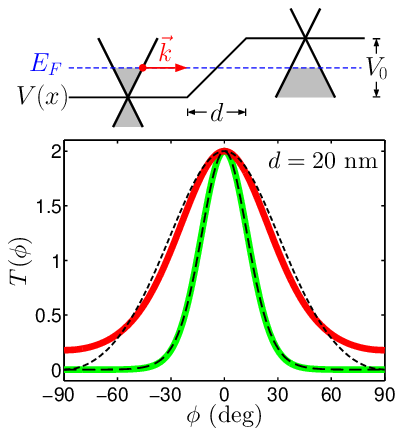}
   \label{fig smooth}}
\caption{(Color online) Klein tunneling in MLG through a \textit{pn}
junction with a (a) sharp and (b) smooth interface. (a) Comparison between
TBM (dashed line) and Eq.\ \eqref{T sharp} \lbrack solid green (gray)]
showing perfect agreement ($E_{F}=80\unit{meV}$). (b) Comparison between TBM
(long and short dashed) and Eq.\ \eqref{T smooth} \lbrack solid green (light
gray) and red (dark gray)] for $k_{F}d\approx 6.16$ ($E_{F}=200\unit{meV}$)
and $k_{F}d\approx 1.54$ ($E_{F}=50\unit{meV}$), respectively.\label{fig sharp vs smooth}}
\end{figure}

\paragraph{Sharp interface}

For a symmetric \textit{pn} junction with a sharp interface [see the
schematic in Fig.\ \ref{fig sharp}], the transmission probability as a
function of $\phi $ was written as\cite{Cheianov2006}%
\begin{equation}
T\left( \phi \right) =\cos ^{2}\phi ,  \label{T sharp}
\end{equation}%
which does not depend on the potential step height. This surprisingly simple
expression matches our TBM result always perfectly as long as $V_{0}=2E_{F}$%
, as shown in Fig.\ \ref{fig sharp}.

For a step potential with arbitrary height $V_{0}\neq 2E_{F}$, the
transmission probability as a function of the incident angle $\phi $ and the
outgoing angle $\theta $ can be derived as%
\begin{equation}
T(\phi ,\theta )=\frac{2\cos \phi \cos \theta }{1+\cos (\phi +\theta )},
\label{T_Dirac}
\end{equation}%
which agrees with our TBM calculation equally well as the symmetric case
(not shown). The two angles $\phi $ and $\theta $ are connected to each
other due to conservation of transverse momentum by%
\begin{equation}
\sin \theta =s\frac{\left\vert E_{F}\right\vert }{\left\vert
E_{F}-V_{0}\right\vert }\sin \phi ,  \label{sin theta}
\end{equation}%
where $s=+1$ for $nn^{\prime }$ or $pp^{\prime }$ and $-1$ for $np$ or $pn$.
Equation \eqref{T_Dirac} clearly recovers the symmetric \textit{pn} junction
case of Eq.\ \eqref{T sharp} when choosing $s=-1$ and $V_{0}=2E_{F}$ in Eq.\
\eqref{sin
theta}. Note that in the case of $\left\vert E_{F}-V_{0}\right\vert
<\left\vert E_{F}\right\vert $, the Fermi wave vector in the outgoing region
is shorter than that in the incoming region, and an additional constraint
for $\phi $ has to be applied to ensure $\left\vert \sin \theta \right\vert
\leq 1$ [i.e., $\phi \leq \left\vert \phi _{c}\right\vert $ with $\phi
_{c}=\sin ^{-1}(\left\vert E_{F}-V_{0}\right\vert /\left\vert
E_{F}\right\vert )$].

Previously it has been stated that the single-valley Dirac picture, based on
which Eqs.\ \eqref{T sharp} and \eqref{T_Dirac} are derived, is not
equivalent to the TBM.\cite{Tang2008} The difference in their work, however, becomes noticeable only when the distance between one of the involved energies and the Dirac point exceeds roughly 300 meV. In our simulation, indeed the deviation for the symmetric {\it pn} junction case with, say $E_F = 300$ meV, is less than $0.5\%$.
The agreement of our TBM and the Dirac theory therefore confirms
that the intervalley scattering, which is mainly responsible for the
nonequivalence at high energies, is indeed negligible.

\paragraph{Smooth interface}

For symmetric \textit{pn} junctions with a linearly varying region of width $%
d$ [see the schematic in Fig.\ \ref{fig smooth}], the analytical derivation
for the transmission probability within the Dirac theory yields\cite%
{Cheianov2006}%
\begin{equation}
T\left( \phi \right) =\exp \left( -\pi \frac{k_{F}d}{2}\sin ^{2}\phi \right)
\label{T smooth}
\end{equation}%
for $k_{F}d\gg 1$.\footnote{%
Note that an additional factor of $1/2$ in the exponent of Eq.\
\eqref{T
smooth} as compared to the original formula given in Ref.\ %
\onlinecite{Cheianov2006} comes from the fact that the linear potential
profile across the interface changes from $-V_{0}$ to $V_{0}$ in Ref.\ %
\onlinecite{Cheianov2006}, but here from $0$ to $V_{0}$, i.e., $k_{F}$
reduces to $k_{F}/2$.} This formula, together with the validity criterion $%
k_{F}d\gg 1$, are tested by our tight-binding calculations shown in Fig.\ %
\ref{fig smooth}, where two sets of parameters are considered. For $%
k_{F}d\approx 6.16$ we find very good agreement with Eq.\ \eqref{T smooth},
while the result for $k_{F}d\approx 1.54$ exhibits noticeable deviations
from the analytical prediction at large angles $\left\vert \phi \right\vert $%
. The smoothing function was assumed in their work as linear but the reality
might be much more complicated, which is then not accessible by the Dirac
theory but again straightforward by our tight-binding calculation.
Nevertheless, the exponential form of Eq.\ \eqref{T smooth} is still a good
description regardless of the actual form of the smoothing function, as we
have numerically checked. What really matters is only the product $k_{F}d$.

Unlike the sharp \textit{pn} interface, a compact form of transmission
probability for the asymmetric case does not exist so far.

\subsection{\textit{pnp} junction: BLG vs MLG+R\label{sec results part 2}}

\begin{figure}[b]
\includegraphics[width=\columnwidth]{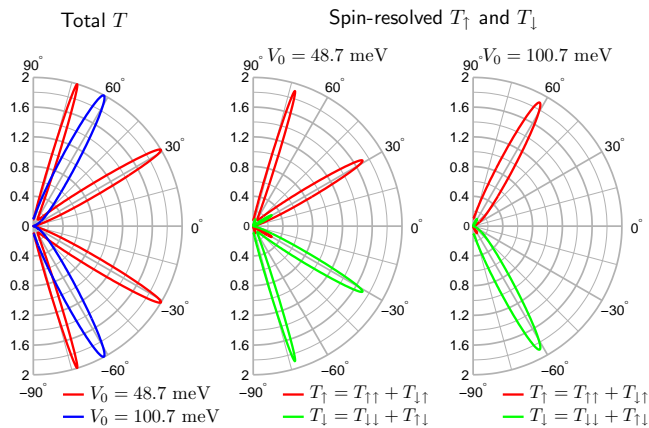}
\caption{(Color online) (a) Angle-resolved total transmission $T$ for
tunneling through a \textit{pnp} junction in MLG+R with the same barrier
height $V_{0}$, barrier width $D$, and Fermi energy $E_{F}$ as used in Fig.\
\protect\ref{fig_pnpBLG_T} for BLG, and a substitution $3t_{R}=\protect%
\gamma _{1}=0.39\unit{eV}$. (b) and (c) show spin-resolved transmission
probabilities for $V_{0}=48.7\unit{meV}$ and $V_{0}=100.7\unit{meV}$,
respectively.\label{fig BLG vs MLGR}}
\end{figure}

We next show the direct correspondence between BLG and MLG+R by considering
exactly the same potential barrier and incident Fermi energy as in Fig.\ \ref%
{fig_pnpBLG_T} for BLG, and set $3t_{R}=\gamma _{1}=0.39\unit{eV}$ for MLG+R
here. (A discussion with weaker, realistic $t_{R}$ will be continued in the
next section.) The total transmission shown in Fig.\ \ref{fig BLG vs MLGR}
for MLG+R indeed resembles the curves in Fig.\ \ref{fig_pnpBLG_T} for BLG,
as expected due to the identical form of their low-energy dispersions %
\eqref{MLG+R} and \eqref{BLG}. The most important feature of chiral
tunneling in BLG, forbidden normal transmission, now appears also in the
case of MLG+R. In BLG, $T\left( \phi =0\right) =0$ was understood as the
consequence of pseudospin conservation. For MLG+R, $T\left( \phi =0\right)
=0 $ can be expected as the consequence of real-spin conservation. Indeed,
this can be demonstrated by computing the nonequilibrium local spin density,
which can be obtained from the lesser Green's function,\cite{Nikolic2006}
considering two cases, $0<E_{F}<3t_{R}$ and $-3t_{R}<E_{F}<0$, both with $%
k_{y}=0$. Within this single-band transmission, the local spin densities for
positive and negative $E_{F}$ point to opposite directions, indicating that
normal incidence transmission between $n$ and $p$ regions will be forbidden.

Next we discuss the spin-resolved transmission. The quantization axis is
chosen as the out-of-plane direction, so that the transmission of, for example, $%
T_{\downarrow \uparrow }$ means the probability of an incoming $+S_{z}$
electron ending up as an outgoing $-S_{z}$ one. Since the \emph{incoming}
angle dependence $\phi $ of the transmission probabilities are analyzed, we
define $T_{\uparrow }=T_{\uparrow \uparrow }+T_{\downarrow \uparrow }$ as
the transmission ability of the $+S_{z}$ electron (or $\uparrow $ spin), and
vice versa. (Alternatively, one can also analyze the \emph{outgoing} angle
dependence and define $T_{\uparrow }$ as $T_{\uparrow \uparrow }+T_{\uparrow
\downarrow }$, not used here. Either way, the total transmission $%
\sum_{\sigma ,\sigma ^{\prime }=\uparrow ,\downarrow }T_{\sigma \sigma
^{\prime }}=T_{\uparrow }+T_{\downarrow }=T$ is ensured.)

The choice of quantization axis $z$ is not necessary but facilitates
relating the present spin-dependent tunneling in MLG with the issue of
intrinsic spin-Hall effect previously discussed in semiconductors. The
spin-resolved transmission curves shown in Fig.\ \ref{fig BLG vs MLGR}
exhibit opposite lateral preference of the $\uparrow $ and $\downarrow $
electron spins, which is an intrinsic spin-Hall mechanism due to the Rashba
SOC. In a\ semiconductor two-dimensional electron gas (i.e., a continuous
system rather than discrete as in the TBM), such an intrinsic spin-Hall
deflection of opposite $S_{z}$ electrons can be easily explained by the
concept of a spin-orbit force based on the Heisenberg equation of motion,%
\cite{Li2005,Nikolic2005}%
\begin{equation}
\mathbf{F}_{so}=\frac{m}{i\hbar }\left[ \frac{1}{i\hbar }\left[ \mathbf{r},%
\mathcal{H}\right] ,\mathcal{H}\right] =\frac{2m\alpha _{R}^{2}}{\hbar ^{3}}%
\left( \mathbf{p}\times \mathbf{e}_{z}\right) \sigma ^{z}.  \label{Fso}
\end{equation}%
Here $\mathcal{H}=p^{2}/2m+(\alpha _{R}/\hbar )(p_{y}\sigma ^{x}-p_{x}\sigma
^{y})$ is the continuous two-dimensional Hamiltonian with Rashba SOC, $%
\mathbf{r}$ and $\mathbf{p}$ are the position and momentum operators, $%
\alpha _{R}$ is the Rashba coupling parameter (rather than the hopping one, $%
t_{R}$), and $\sigma ^{z}$ is the sign of the $S_{z}$ spin component. The $%
T_{\uparrow }$ and $T_{\downarrow }$ curves shown in Fig.\ \ref{fig BLG vs
MLGR} therefore reveal a combined effect of forbidden normal transmission
due to conservation of real spin and the intrinsic spin-Hall deflection that
can be understood by Eq.\ \eqref{Fso}.

A few remarks are due before we move on. To connect BLG with MLG+R we put $%
3t_{R}=\gamma _{1}=0.39\unit{eV}$, which is apparently far from reality. In
general the Rashba splitting induced by electrical gating is roughly of or
less than the order of $100%
\unit{\mu eV}%
$ (see Sec.\ \ref{sec exp}). Fermi energy\ lying within this splitting,
which is also our main interest, projects to a much shorter Fermi wave
vector $k_{F}$, leading to a much longer $d$ up to a few or a few tens of
microns in order for $k_{F}d\gg 1$ to be valid. This implies that the
influence of the interface on the tunneling in MLG+R is normally negligible,
unless $d$ is that long. In addition, tunneling through a \textit{pnp}
junction will also require a long barrier width $D$ for electrons subject to
such a short $k_{F}$; otherwise, the barrier is merely a weak perturbation
to the electron due to its long Fermi wave length. Based on these remarks,
we will focus in the next section only on \textit{pn} junctions in MLG+R with a
reasonable Rashba hopping parameter.

\begin{figure}[t]
\includegraphics[width=\columnwidth]{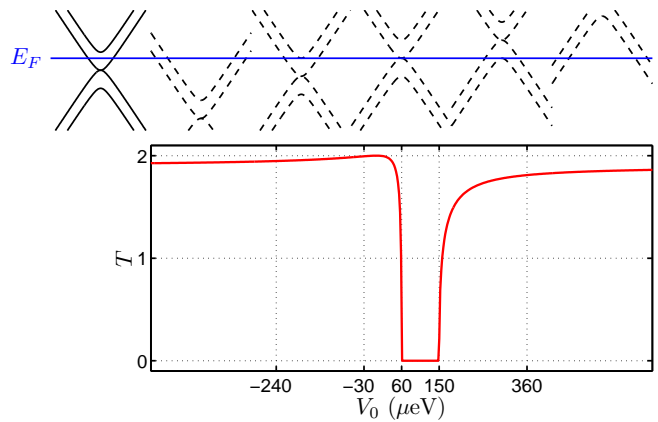}
\caption{(Color online) Transmission $T$ at normal incidence ($k_{y}=0$) as
a function of potential step height $V_{0}$ for tunneling through a \textit{%
pn} junction in MLG+R. The leftmost solid band diagram above the main panel
corresponds to the incoming \textit{n} side. The five ticks on the $V_{0}$
axis correspond to the above five dashed band diagrams for the outgoing
side.\label{fig T normal}}
\end{figure}

\subsection{\textit{pn} junction in MLG+R\label{sec results part 3}}

In the following we demonstrate in detail the role of Rashba SOC in
tunneling through a potential step in MLG+R. The Rashba hopping parameter
will be fixed to $t_{R}=30%
\unit{\mu eV}%
$ and the Fermi energy in most cases to $E_{F}=2t_{R}$, which lies within
the spin-orbit splitting $3t_{R}$ (see Fig.\ \ref{fig1}).

\subsubsection{Normal incidence}

We begin with the case of normal incidence, $k_{y}=0$. In Sec.\ \ref{sec
results part 2} we have discussed the one-band transmission selection rule
(i.e., $n\leftrightarrow p$ transmission is forbidden). The transmission from
the left side at Fermi energy $0<E_{F}<3t_{R}$ to the right side with
potential $V_{0}$ is expected to be zero whenever a single-band $%
n\rightarrow p$ transmission is attempted. Indeed, as shown in Fig.\ \ref%
{fig T normal}, a zero transmission gap of $T$ as a function of $V_{0}$ is
found. The gap lies in the interval of $E_{F}<V_{0}<E_{F}+3t_{R}$,
corresponding to the single-band $n\rightarrow p$ transmission. Note that
contrary to the valley-valve effect in zigzag nanoribbons,\cite%
{Wakabayashi2002,Rycerz2007,Cresti2008} the gap shown here arises solely due
to a bulk property.

\begin{figure}[b]

\includegraphics[width=\columnwidth]{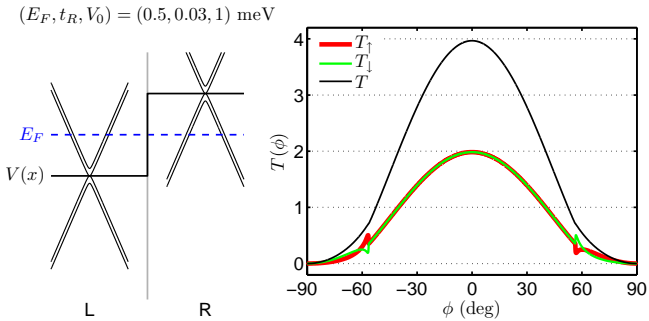}
\caption{(Color online) Angular dependence of total ($T$) and spin-resolved (%
$T_{\uparrow }$ and $T_{\downarrow }$) transmissions for $E_{F}=0.5\unit{meV}
$ well above the Rashba splitting $3t_{R}=90%
\unit{\mu eV}%
$.\label{fig T vs phi trivial}}
\end{figure}

\subsubsection{Angle- and spin-resolved transmission}

We proceed with angle- and spin-resolved transmission and consider first the
trivial case with $E_{F}=0.5\unit{meV}$ well above the Rashba splitting $%
3t_{R}=90%
\unit{\mu eV}%
$, as shown in Fig.\ \ref{fig T vs phi trivial}. In this case the maximum of
$T=T_{\uparrow }+T_{\downarrow }$ is $4$ since two spin subbands and two
valleys are involved in transport. The total transmission curve resembles
the expected $\cos ^{2}\phi $ behavior as discussed in Sec.\ \ref{sec falko}%
, showing that the Rashba effect plays only a minor role. The spin-resolved $%
T_{\uparrow }$ and $T_{\downarrow }$ curves differ only slightly at $%
\left\vert \phi \right\vert =\sin ^{-1}(k_{F}^{\text{in}}/k_{F}^{\text{out}%
})\approx 56\unit{%
{{}^\circ}%
}$, where $k_{F}^{\text{in}}$ and $k_{F}^{\text{out}}$ are the inner and
outer radius of the two concentric Fermi circles, respectively. Tunneling in
BLG with $E_{F}$ well above $\gamma _{1}$ behaves similarly (i.e., the
interlayer coupling $\gamma _{1}$ in BLG no longer plays an important role
in the process of chiral tunneling when the transport occurs at $E_{F}\gg
\gamma _{1}$), as we have numerically checked. In other words, the chiral
tunneling in BLG with $E_{F}\gg \gamma _{1}$ and in MLG+R with $E_{F}\gg
3t_{R}$ recovers the Klein tunneling behavior as in MLG.

\begin{figure}[t]
\subfigure[]{
   \includegraphics[width=\columnwidth]{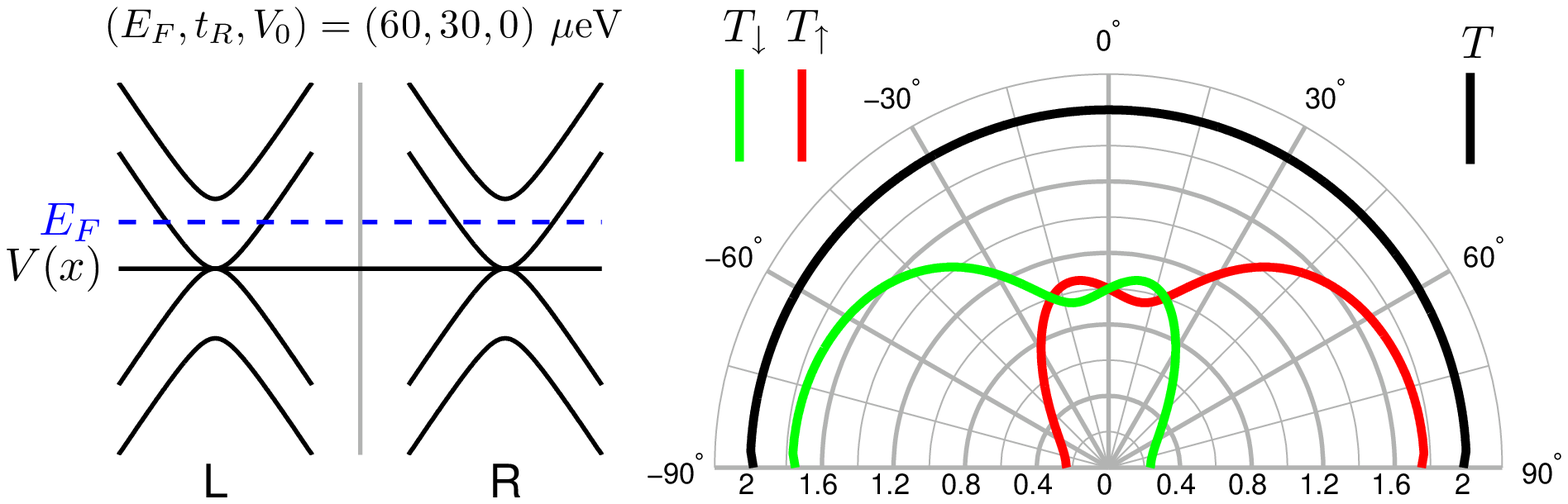}
   \label{fig T oblique 1}}
\par
\subfigure[]{
   \includegraphics[width=\columnwidth]{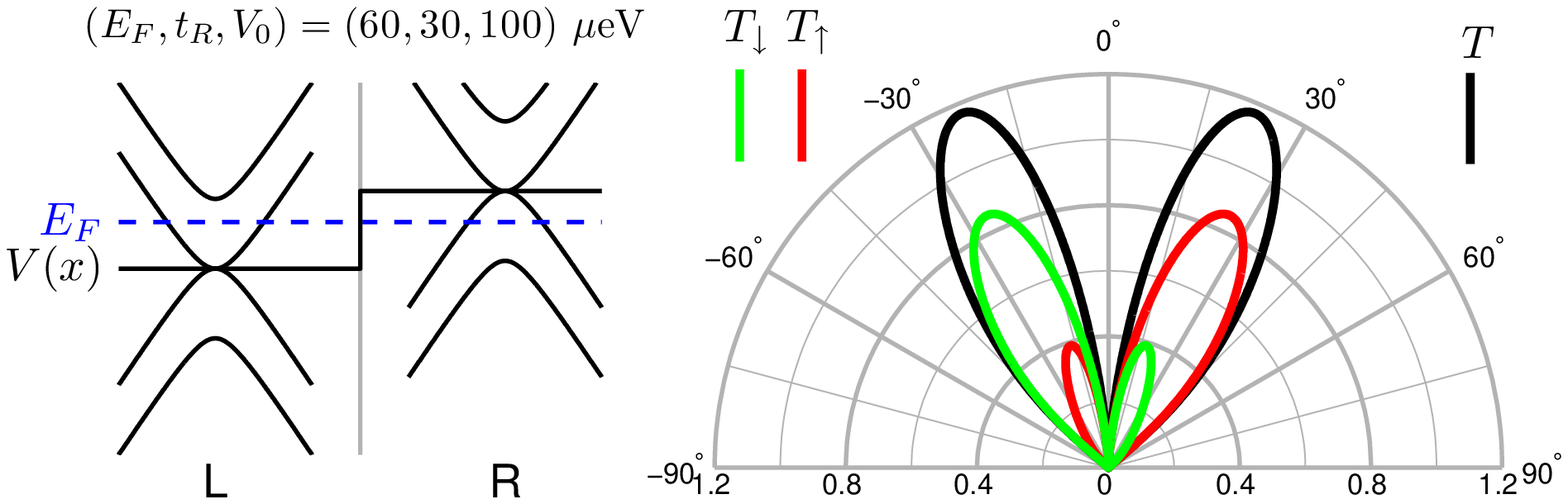}
   \label{fig T oblique 2}}
\caption{(Color online) Angular dependence of total and spin-resolved
transmissions through a \textit{pn} junction in MLG+R with (a) zero
potential and (b) finite potential. Parameters used are given above.\label{fig T oblique}}
\end{figure}

Of particular interest is the nontrivial case with $\left\vert
E_{F}\right\vert <3t_{R}$. As a test, we first consider $V_{0}=0$ as shown
in Fig.\ \ref{fig T oblique 1}. In the absence of the potential step, the
total transmission function $T$ reaches its maximum of $2$ (one spin subband
times valley degeneracy of two) for any angle $\phi $, as it should. The
opposite lateral deflection tendency of the $\uparrow $ and $\downarrow $
spins is again clearly seen and can be explained based on Eq.\ \eqref{Fso}
as discussed in Sec.\ \ref{sec results part 2}.

The most important case is that of Fermi energy $E_{F}\in \left(
0,3t_{R}\right) $ and potential height $V_{0}\in \left(
E_{F},E_{F}+3t_{R}\right) $. A specific example with $V_{0}=100%
\unit{\mu eV}%
$ is shown in Fig.\ \ref{fig T oblique 2}, which exhibits the combined
effect of the forbidden normal transmission [$T(\phi =0)=0$] and spin-Hall
deflection. The number of high transmission peaks is always two.\footnote{%
We have numerically checked that the double-peak feature of $T_{\text{tot}}$
in the single-band $n\leftrightarrow p$ transmission regime shown in Fig.\ %
\ref{fig T oblique 2} still holds even if the intrinsic SOC is present, as
long as the Rashba coupling dominates.} Compared to the previous trivial
case ($E_{F}>3t_{R}$, Fig.\ \ref{fig T vs phi trivial}) where $T_{\uparrow }$
and $T_{\downarrow }$ do not significantly differ, the separation of the
opposite $\uparrow $ and $\downarrow $ spins is distinctly enhanced. Whether
this could be a new type of intrinsic spin-Hall mechanism in graphene
deserves a further investigation, and is left as a possible future
direction.
\begin{figure*}[t]
\includegraphics[width=0.9\textwidth]{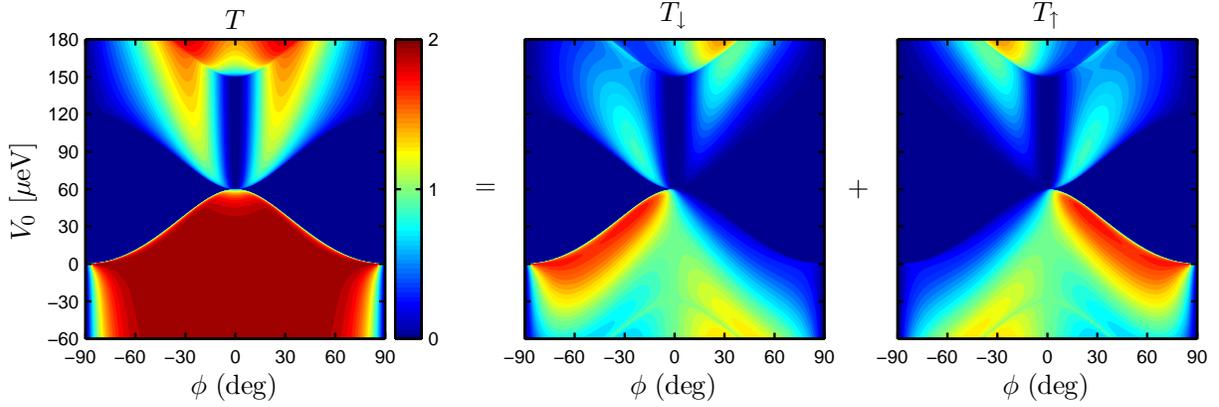}
\caption{(Color online) Transmission through a \textit{pn} junction in MLG+R
as a function of incident angle $\protect\phi $ and potential step height $%
V_{0}$. Four transmission regimes can be distinguished: (i) $V_{0}<0$, (ii) $%
0<V_{0}<E_{F}$, (iii) $E_{F}<V_{0}<E_{F}+3t_{R}$, and (iv) $%
V_{0}>E_{F}+3t_{R}$, with $E_{F}=60%
\unit{\mu eV}%
$ and $3t_{R}=90%
\unit{\mu eV}%
$.\label{fig T vs phi vs V0}}
\end{figure*}

We summarize the discussion of angle- and spin-resolved transmission by
mapping $T\left( \phi ,V_{0}\right) $ in Fig.\ \ref{fig T vs phi vs V0}.
Four different transport regimes can be identified:

\begin{enumerate}
\item $V_{0}<0$, single $n$ band to single/multiple $n$ band(s) transmission
regime.

\item $0<V_{0}<E_{F}$, single $n$ band to single $n$ band transmission
regime; distinct spin-resolved $T_{\uparrow }$ and $T_{\downarrow }$, and
high total $T$ limited by a critical angle $\phi _{c}=\sin ^{-1}(\left\vert
E_{F}-V_{0}\right\vert /\left\vert E_{F}\right\vert )$.

\item $E_{F}<V_{0}<E_{F}+3t_{R}$, single $n$ band to single $p$ band
transmission regime; combined effects of forbidden normal transmission and
spin-Hall deflection.

\item $V_{0}>E_{F}+3t_{R}$, single $n$ band to multiple $p$ bands
transmission regime.
\end{enumerate}

Note that a vertical scan in Fig.\ \ref{fig T vs phi vs V0} at $\phi =0$
corresponds to Fig.\ \ref{fig T normal}, and horizontal scans at $V_{0}=0$
and $V_{0}=100%
\unit{\mu eV}%
$ to Figs.\ \ref{fig T oblique 1} and \ref{fig T oblique 2}, respectively.
These four regimes will be helpful in the following discussion of
conductance.

\begin{figure}[b]
\includegraphics[width=\columnwidth]{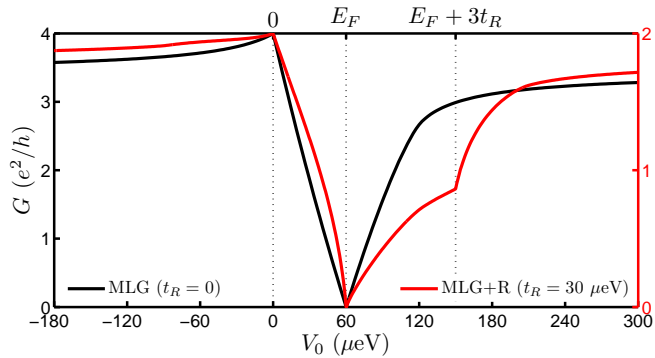}
\caption{(Color online) Integrated conductance of the ballistic \textit{pn}
junction in MLG with $t_{R}=0$ and MLG+R with $t_{R}=30\unit{\protect\mu eV}$%
.\label{fig G}}
\end{figure}

\subsubsection{Integrated conductance}

Finally, we calculate the conductance of the \textit{pn }junction in MLG+R
by integrating $T\left( \phi \right) $, or equivalently, $T\left(
k_{y}\right) $, with respect to the transverse Bloch momentum,%
\begin{equation}
G=\frac{e^{2}/h}{2k_{F}}\int_{-k_{F}}^{k_{F}}T\left( k_{y}\right) dk_{y},
\label{G}
\end{equation}%
where the prefactor ensures the maximal value of the Landauer-B\"{u}ttiker
type ballistic conductance to be $e^{2}/h$ times the maximal number of modes.%
\cite{Datta1995} We compare the conductance of the \textit{pn} junction in
MLG ($t_{R}=0$) and in MLG+R ($t_{R}=30%
\unit{\mu eV}%
$) as a function of the potential step height $V_{0}$, as shown in Fig.\ \ref%
{fig G}. Since the Fermi level is fixed to $E_{F}=60%
\unit{\mu eV}%
$ for both cases, the transport for $t_{R}=0$ will involve two spin and two
valley degeneracies, leading to the maximal $G$ of $4e^{2}/h$, while in the
case of $t_{R}=30%
\unit{\mu eV}%
$ only one spin subband is projected, leading to the maximal $G$ of $%
2e^{2}/h $. The maximal $G$ occurs always at $V_{0}=0$ that corresponds to
an ungated clean bulk graphene. Zero conductance, on the other hand, occurs
at $V_{0}=E_{F}$ since no states at the outgoing region are available at
this charge neutrality point.

Different transmission regimes can be distinguished based on our previous
discussion for Fig.\ \ref{fig T vs phi vs V0}. For $V_{0}\in \lbrack 0,60]%
\unit{\mu eV}%
$ ($n\rightarrow n$ transmission), the rise of $V_{0}$ shrinks the Fermi
circle at the outgoing region and hence introduces a critical transverse
momentum, outside which the transmission is suppressed due to the lack of
out-going states. The critical transverse momentum reduces linearly with $%
V_{0}$ for MLG due to the linear dispersion. The conductance $G$, Eq.\ %
\eqref{G}, therefore reduces also linearly with $V_{0}$. In the presence of
the Rashba SOC, the low-energy dispersion becomes quadratic, and so does the
reduction of $G$ with $V_{0}$ in MLG+R.

For $V_{0}\in \lbrack 60,150]%
\unit{\mu eV}%
$ ($n\rightarrow p$ transmission), the conductance of MLG rises faster than
that of MLG+R, possibly due to the help of Klein tunneling. At $V_{0}=150%
\unit{\mu eV}%
$, a sudden jump (or a shoulder) occurs in the case of MLG+R since the
second spin subband at the outgoing region starts to participate in
transport. This jump does not occur in the MLG case since both spin subbands
are always degenerate. An earlier related work based on Dirac theory
considered both intrinsic and Rashba SOCs.\cite{Yamakage2009} The $V_{0}$
dependence of $G$ for the Rashba dominated case in that work agrees well
with the MLG+R curve shown in Fig.\ \ref{fig G}, including the shoulder.

\section{Experimental aspects\label{sec exp}}

\subsection{Rashba spin splitting in graphene}

Whereas the Rashba spin splitting in MLG induced by an applied electric
field is in general in the order of no more than $100%
\unit{\mu eV}%
$, which is beyond the present resolution of angle-resolved photoelectron
spectroscopy (ARPES), direct experimental observation of the Rashba spin
splitting at $K$ and $K^{\prime }$ in agreement with the first-principles
calculations\cite{Gmitra2009,Abdelouahed2010} is so far not reported. An
earlier experiment on epitaxial graphene layers on a Ni(111) surface
reported a large Rashba interaction\cite{Dedkov2008} up to $225\unit{meV}$
but was soon questioned since the splitting might simply reveal a Zeeman
type splitting due to the ferromagnetic nature of nickel.\cite{Rader2009} An
intercalated Au monolayer between the graphene layer and the Ni(111)
substrate reduced the splitting to about $13\unit{meV}$ and was concluded as
the Rashba effect on the $\pi $ states supported by spin-resolved ARPES.\cite%
{Varykhalov2008} However, the low-energy band structure of MLG+R at that
time was not yet clear, and a simplified picture was adopted in the
explanation of the measured spin splitting. In addition, transport
properties of graphene based on metallic substrates can be difficult to
isolate since a large bulk current will interfere as background.\cite%
{Yaji2010}

Throughout the above calculations we have mostly focused on a rather weak
Rashba hopping parameter $t_{R}=30%
\unit{\mu eV}%
,$ yielding a splitting at the $K$ and $K^{\prime }$ points $3t_{R}=90%
\unit{\mu eV}%
$, which is a realistic and rather conservative estimate for the
gate-voltage-induced Rashba SOC strength. A recent proposal of
impurity-induced SOC in graphene,\cite{CastroNeto2009a} however, indicated
that the coupling strength can be strongly enhanced by putting heavy adatoms%
\cite{Weeks2011} as well as by hydrogenation.\cite{CastroNeto2009a,Elias2009}

\subsection{Klein tunneling in MLG}

Indirect and direct experimental evidences of Klein tunneling in MLG have
been reported recently.\cite{Stander2009,Young2009}. For detailed reviews,
we refer to Refs.\ %
\onlinecite{Beenakker2008,CastroNeto2009,DasSarma2011,Young2011,Allain2011}.
A very recent experiment on transport through a \textit{pnp} junction in MLG
used an embedded local gate, which yields high quality ballistic transport
and perfectly independent control of the local carrier density, as well as
the feature of Klein tunneling.\cite{Nam2011}

Recall the $t_{R}=0$ curve of conductance for MLG shown in Fig.\ \ref{fig G}%
. Overall, the conductance for $n\rightarrow n$ transmission with $V_{0}<0$
is always higher than that for $n\rightarrow p$ transmission with $%
V_{0}>E_{F}$. Even though Klein tunneling leads to perfect transmission at
normal incidence in the latter case, the decay of $T$ with incident angle
eventually yields a lower conductance after integration. This feature has
been agreed in recent experiments for \textit{pn} and \textit{pnp }junctions
in MLG.\cite%
{Huard2007,Williams2007,Ozyilmaz2007,Liu2008c,Stander2009,Young2009,Nam2011,Gabor2011}
The difference of the conductance, or equivalently the resistance, between
the \textit{nn }and\textit{\ np} (or between \textit{pp }and\textit{\ pn})
in experiments is even more obvious possibly due to the smooth interface
that leads to an exponentially decaying form of $T$,\cite{Cheianov2006} as
we have reviewed and discussed in Sec.\ \ref{sec falko}. In fact, for MLG we
have numerically checked $G$ for \textit{pn} junctions with a smooth
interface, which indeed can enhance the difference of $G$ between the
\textit{nn }and\textit{\ np} regimes.

Another interesting feature so far experimentally reported only in Refs.\ %
\onlinecite{Young2009} and \onlinecite{Nam2011} is the Fabry-Perot
oscillation of the conductance for \textit{pnp }junctions due to the
interference between the two interfaces of the central barrier. This feature
requires the system to be ballistic and can be naturally revealed by our
tight-binding transport calculation, which we will elaborate elsewhere in
the future.

\section{Conclusion and outlook\label{sec conclusion}}

In conclusion, we have employed tight-binding calculations to show that
transport properties of MLG+R behave as BLG due to their identical form of
the low-energy dispersion, choosing the chiral tunneling in \textit{pn} and
\textit{pnp} junctions as a concrete example. Within single-band
transmission, normal incidence transmission through a \textit{pn} junction
in BLG with $\left\vert E_{F}\right\vert <\gamma _{1}$ is forbidden as a
consequence of pseudospin conservation,\cite{Katsnelson2006} while in MLG+R
with $\left\vert E_{F}\right\vert <3t_{R}$ this forbidden transmission also
occurs but as a consequence of real-spin conservation. In mapping the angle-
and spin-resolved transmission for the MLG+R case, a combined effect of
forbidden normal transmission and intrinsic spin-Hall deflection is revealed
[Fig.\ \ref{fig T oblique 2}]. Compared to the potential-free spin-Hall
deflection case as shown in Fig.\ \ref{fig T oblique 1}, where $T_{\uparrow
}=T_{\downarrow }=1$ at $\phi =0$, the effect of the \textit{pn} junction
seems to force the up and down spins to separate since $T_{\uparrow
}=T_{\downarrow }=0$ at $\phi =0$. The feature revealed in Fig.\ \ref{fig T
oblique 2} may therefore suggest a new type of intrinsic spin-Hall mechanism
in MLG.

Within multiband transmission, however, the Rashba SOC in MLG no longer
plays an important role when $\left\vert E_{F}\right\vert \gg 3t_{R}$ (Fig.\ %
\ref{fig T vs phi trivial}). Likewise, the interlayer hopping $\gamma _{1}$
in BLG becomes unimportant when $\left\vert E_{F}\right\vert \gg \gamma _{1}$%
. Transport in both MLG+R with $\left\vert E_{F}\right\vert \gg 3t_{R}$ and
BLG with $\left\vert E_{F}\right\vert \gg \gamma _{1}$ recovers to that in
MLG, despite the usually very different energy scales of $3t_{R}$ and $%
\gamma _{1}$. In view of the distinct transmission patterns in MLG+R with $%
\left\vert E_{F}\right\vert <3t_{R}$ [Fig.\ \ref{fig T oblique 2}] and $%
\left\vert E_{F}\right\vert \gg 3t_{R}$ (Fig.\ \ref{fig T vs phi trivial}),
as an interesting conjecture for the BLG case one expects very different
scattering regimes for $\left\vert E_{F}\right\vert <\gamma _{1}$ and $%
\left\vert E_{F}\right\vert \gg \gamma _{1}$. The former is well discussed
in the literature and exhibits strong scattering [Fig.\ \ref{fig_pnpBLG_T}]
while the latter is less discussed and the scattering is expected to be
strongly suppressed.

MLG and BLG are known to behave quite differently in general, in the sense
of single-band transmission. Whereas turning MLG directly into BLG is in
principle not possible, steering MLG to MLG+R can be achieved simply by
gating, and therefore the effect of Rashba SOC provides a possibility to
continuously change the MLG-like transport properties to BLG-like. We expect
further transport properties to behave similarly in BLG and in MLG+R, such
as the quantum Hall effect,\cite{Novoselov2005} as was also noted by Rashba.%
\cite{Rashba2009}

\begin{acknowledgments}
We\ gratefully acknowledge Alexander von Humboldt Foundation (M.H.L.) and
Deutsche Forschungsgemeinschaft (within SFB689) (J.B. and K.R.) for
financial support.
\end{acknowledgments}

\bibliographystyle{apsrev4-1}
\bibliography{mhl2}

\end{document}